\documentstyle[aps,prd,times]{revtex}
\input{epsf}
\begin{document}

\draft

\title{Cauchy-perturbative matching and outer
       boundary conditions: computational studies}

\author{Luciano~Rezzolla$^{\rm a,b}$, Andrew~M.~Abrahams$^{\rm a,c}$, 
        Richard~A.~Matzner$^{\rm d}$, Mark~E.~Rupright$^{\rm e}$ 
        and Stuart~L.~Shapiro$^{\rm a,f,b}$}

\address{$^{\rm a}$Department of Physics, University of Illinois at
        Urbana-Champaign, Urbana, Illinois 61801}
\address{$^{\rm b}$NCSA, University of Illinois at
        Urbana-Champaign, Urbana, Illinois 61801}
\address{$^{\rm c}$J.~P. Morgan, 60 Wall St., New York, New York 10260}
\address{$^{\rm d}$The University of Texas at Austin, Austin, 
        Texas, 78712}
\address{$^{\rm e}$Department of Physics and Astronomy, University of
         North Carolina, Chapel Hill, North Carolina 27599-3255}
\address{$^{\rm f}$Department of Astronomy, University of Illinois at
        Urbana-Champaign, Urbana, Illinois 61801}

\date{\today}
\maketitle

\begin{abstract}
	We present results from a new technique which allows
extraction of gravitational radiation information from a generic
three-dimensional numerical relativity code and provides stable outer
boundary conditions. In our approach we match the solution of a Cauchy
evolution of the nonlinear Einstein field equations to a set of
one-dimensional linear equations obtained through perturbation
techniques over a curved background. We discuss the validity of this
approach in the case of linear and mildly nonlinear gravitational
waves and show how a numerical module developed for this purpose is
able to provide an accurate and numerically convergent description of
the gravitational wave propagation and a stable numerical evolution.
\end{abstract}


\section{Introduction}
\label{intro}

	In the past few years a considerable effort has been devoted
to the solution of Einstein's equations in numerical simulations of
strong-field, highly dynamical sources of gravitational
radiation. This effort is partly motivated by the development and
construction of gravitational wave detectors such as LIGO, VIRGO, GEO
and TAMA. Knowing the theoretical waveform produced by the most likely
astrophysical sources of gravitational radiation will not only
increase the probability of a successful detection but, most
importantly, will allow for the extraction of astrophysically
significant information from the observations.
	
	The Binary Black Hole ``Grand Challenge'' {\it Alliance}
\cite{BBHGCA}, is a major example of this effort, in which a
multi-institutional collaboration in the United States was created in
order to study the inspiral and coalescence of a binary black hole
system, one of the most significant source of signals for the
interferometric gravity wave detectors. Present three-dimensional (3D)
numerical relativity simulations face a number of fundamental and in
some cases unsolved problems, including: coordinate choice, most
suitable form of Einstein's equations, singularity avoiding
techniques, gravitational wave extraction and outer boundary
conditions. While a robust solution to the generic problem is still
awaited, some interesting results have already been obtained, for
instance, in the evolution of a generic 3D black hole \cite{prl3}, in
the translation of a 3D black hole across a numerical grid
\cite{prl2}, and in the extraction of gravitational wave information
and imposition of outer boundary conditions \cite{prl1}.

	Determining the asymptotic form of the gravitational waves
produced in a dynamical evolution of Einstein's equations is an
important goal of many numerical relativity simulations. This goal,
however, necessarily requires accurate techniques which compute
waveforms from numerical relativity simulations on 3D spacelike
hypersurfaces with finite extents. In an ideal situation with
unlimited numerical resources, the computational domain can extend
into the distant wave zone \cite{Thorne}, where the geometric optics
approximation is valid and the gravitational waves approach their
asymptotic form.  With present-day computational limitations, however,
the outer boundary of typical numerical relativity simulations lies
rather close to the highly dynamical and strong-field region where
backscatter of waves off curvature can be significant.  As a result,
additional techniques need to be implemented in order to ``extract''
such information from the strong-field region and ``evolve'' it out to
a large distance.

	In two recent papers \cite{prl1,prd1}, we have presented a new
method for extracting gravitational wave data from a 3D numerical
relativity simulation and evolving it out to an arbitrary distant
zone. Our method has been developed within the Alliance in order to
match a generic full 3D Cauchy solution of nonlinear Einstein's
equations on spacelike hypersurfaces with a linear solution in a
region where the waveforms can be treated as perturbations on a
spherically symmetric curved background. This ``perturbative module''
is used not only to extract gravitational wave data from the Cauchy
evolution but, at the same time, to impose outer boundary conditions
(A parallel development is also underway in the Alliance to match
interior Cauchy solutions to exterior solutions on characteristic
hypersurfaces \cite{characteristic}). Indeed, while the problem of
radiation extraction is important for computing observable waveforms
from numerical simulations, imposition of correct outer boundary
conditions is essential for maintaining the integrity of the
simulations themselves, as incorrect outer boundary conditions are
often a likely source of numerical instabilities. One of the most
important requirements for any radiation-extraction and outer-boundary
module is that it provides for stable evolution of the interior
equations and minimizes the spurious (numerical) reflection of
radiation at the boundary. This requirement is particularly important
for the ``Grand Challenge'' investigation in which the computations
need to be performed on a gravitational wave emission timescale, which
is much longer than the orbital one.

	 In this paper we present the application of our
Cauchy-perturbative matching method to a standard testbed: the
evolution of 3D linear and mildly nonlinear gravitational waves. The
plan of this paper is as follows: in Section \ref{tpm} we briefly
review the main features of the approach and recall the essential
elements of its numerical implementation. We then concentrate on the
two major aspects of this work: in Section \ref{stb} we present the
{\it ``short term''} properties of the Cauchy-perturbative matching
and show its ability to provide an accurate and numerically convergent
approximation to the gravitational waveform that would be observed in
the wave zone surrounding an isolated source. In Section \ref{ltb} we
turn to the {\it ``long term''} properties of our approach and present
a number of different implementations which lead to a {\it stable}
numerical evolution, long after the bulk of the gravitational waves
have left the computational grid.

\section{The Perturbative Method}
\label{tpm}

	As discussed in \cite{prd1} (hereafter Paper I), the
Cauchy-perturbative matching method involves replacing, at least in
parts of the 3D numerical domain, the solution of the full nonlinear
Einstein's equations with the solution of a set of simpler linear
equations that can be integrated to high accuracy with minimal
computational cost.

\begin{figure}[h]
\epsfxsize=10.0cm
\begin{center}
\vskip 1.0truecm 
\leavevmode
\epsffile{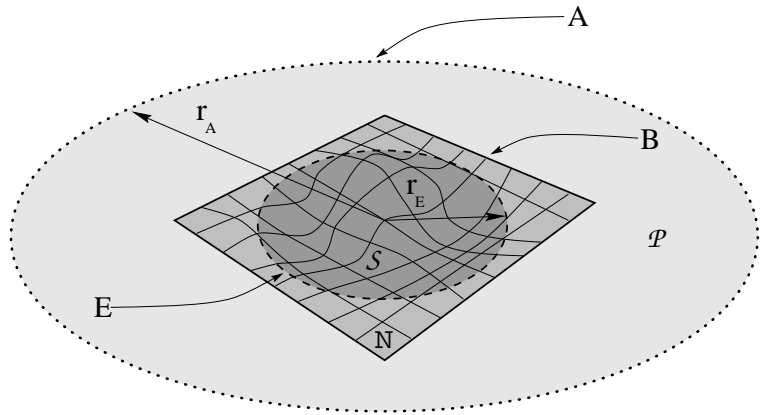}
\vskip 1.0truecm 
\end{center}
\caption[fig1]{\label{fig1}
	Schematic picture of the Cauchy-perturbative matching
	procedure for a spacelike slice of spacetime (one dimension
	has been suppressed). {\tt N} is the 3D numerical grid in
	which the full Einstein's equations are solved and {\bf B} its
	2D outer boundary. The interior (dark shaded) region ${\cal
	S}$ shows the strong-field highly dynamical region of
	spacetime fully covered by {\tt N}. ${\cal P}$ is the region
	of spacetime where a perturbative solution can be performed
	and extends from the 2-sphere {\bf E} (of radius $r_{_E}$) to
	the 2-sphere {\bf A} (of radius $r_{_A}$) located in the
	asymptotically flat region of spacetime. ${\cal P}$ is covered
	entirely by a 1D grid {\tt L} (not shown) and partially by the
	3D grid {\tt N}.}
\end{figure}

	In order to do this, it is necessary to determine the region
of spacetime where a perturbative approach can be applied. In general,
the 3D numerical grid (indicated as {\tt N} in \hbox{Fig. \ref{fig1}})
will comprise an isolated region of spacetime where the gravitational
fields are strong and highly dynamical.  In this region, indicated as
${\cal S}$ in \hbox{Fig. \ref{fig1}}, the full nonlinear Einstein
equations must be solved.  Outside of ${\cal S}$, however, in what we
will refer to as the perturbative region ${\cal P}$, a perturbative
approach is not only possible but highly advantageous.  Anywhere in
the portion of ${\cal P}$ covered by {\tt N} we can place a
two-dimensional (2D) surface which will serve as the surface joining
numerically the highly dynamical strong-field region ${\cal S}$ and
the perturbative one ${\cal P}$. Here, we have chosen this surface to
be a 2-sphere of radius $r_{_E}$, indicated as {\bf E} in
\hbox{Fig. \ref{fig1}}. It is important to emphasize that the 2-sphere
{\bf E} {\it need not} be in a region of spacetime where the
gravitational fields are {\it weak} or the curvature is {\it
small}. In contrast to previous investigations which matched
Einstein's equations onto a Minkowski background \cite{ae}, the
matching is here made on a Schwarzschild background. As a result, the
only requirement is that the spacetime outside of ${\cal S}$
approaches a Schwarzschild one. Even in the case of a binary black
hole merger, it will be possible to find a region of spacetime,
sufficiently distant from the binary black holes, where this
requirement is met \cite{pp94,ac94,ast95}.

	In a practical implementation of the Cauchy-perturbative
method, a numerical code provides the solution to the full nonlinear
Einstein equations everywhere in the 3D grid {\tt N} except at
its outer boundary surface {\bf B}. At the extraction 2-sphere {\bf
E}, a different code (i.e. the perturbative module) ``extracts'' the
gravitational wave information and transforms it into a set of
multipole amplitudes which we have here chosen to depend only on the
radial and time coordinates of the background Schwarzschild metric
(see Paper I and Section \ref{tbe} for details)\footnote{Note that
		although highly convenient, the suppression of the
		angular part of the multipoles is not strictly
		necessary. Indeed, different linear perturbation
		equations can be derived in which the angular
		dependence is explicitly contained in the evolution 
		equations.}.

	In this way, two of the three spatial dimensions of the
problem are suppressed and the propagation of gravitational waves on a
curved background is reduced to a one-dimensional (1D) problem. During
each timestep, information about the gravitational field read-off at
{\bf E} is propagated by the perturbative module out to the 2-sphere
{\bf A} in the asymptotic flat region of spacetime. This is done by
solving a set of coupled 1D linear differential equations (one for
each of the multipoles extracted at {\bf E}) on the 1D grid {\tt L}
(not shown in \hbox{Fig. \ref{fig1}}) covering the perturbative region
${\cal P}$ and ranging between $r_{_E}$ and $r_{_A} \gg r_{_E}$. From
a computational point of view, this represents an enormous advantage:
with a few straightforward transformations, the computationally
expensive 3D evolution of the gravitational waves via the nonlinear
Einstein equations is replaced with a set of 1D linear equations that
can be integrated to high accuracy with minimal computational
cost. Although linear, these equations account for all of the effects
of wave propagation in a curved spacetime and, in particular,
automatically incorporate the effects of backscatter off the
curvature (only the wave-wave effects are omitted).

	As a result of our construction (and as shown in
\hbox{Fig. \ref{fig1}}), the perturbative region ${\cal P}$ is
entirely covered by a 1D grid {\tt L} and only partially by a 3D grid
in the complement to ${\cal S}$ in {\tt N}. The overlap between these
two grids is essential. In fact, the knowledge of the solution on
${\cal P}$ allows the perturbative module to provide boundary
conditions at the outer boundary surface {\bf B} and, if useful,
Dirichlet data on every gridpoint of {\tt N} outside the strong region
${\cal S}$. As we will further discuss in Section \ref{pb}, this
freedom to specify boundary data on a 2-surface of arbitrary shape as
well as on a whole 3D region of {\tt N} represents an important
advantage of the perturbative module over similar approaches to the
problem of gravitational wave extraction and imposition of boundary
conditions.

\subsection{The basic equations}
\label{tbe}

	Our treatment of Schwarzschild perturbation theory is based on
the third-order Einstein-Ricci hyperbolic formulation of Einstein
field equations \cite{cby95,aacby95}. A principal advantage of this
approach is that gauge-invariant wave equations arise simply from a
linear reduction of the full equations without complex changes of
variables. Thus, the matching of the perturbative solutions to the
fully nonlinear ones becomes rather straightforward. Once the
perturbative equations are derived, these are completely general and
can be applied to numerical codes solving Einstein's equations in
either an explicitly hyperbolic form or, as in the present case, in
the standard $3+1$ form~\cite{York79}.

	 We split the gravitational quantities of interest into
background parts (denoted by a tilde) and perturbed parts. These are
the three-metric $g_{ij} = \tilde g_{ij} + h_{ij}$, the extrinsic
curvature $K_{ij} = \tilde{K}_{ij} + \kappa_{ij}$, the lapse function
$N = \tilde{N} + \alpha$ and the shift vector $\beta^i = \tilde
{\beta}^i + v^i$, where the tilde denotes background
quantities. Assuming a Schwarzschild background,

\begin{eqnarray}
\tilde{g}_{ij} dx^i dx^j &=& \tilde{N}^{-2} dr^2 + 
	r^2 (d\theta^2 + \sin^2\theta d\phi^2) \ , \\
\tilde{N} & = & \left(1-\frac{2M}{r}\right)^{1/2} \ , 
\end{eqnarray}
we then have $\tilde{K}_{ij}=0=\tilde {\beta}^i$, while the perturbed
parts have arbitrary angular dependence.

	Using this background, we linearize the hyperbolic equations
and reduce the wave equation for $K_{ij}$ to a linear wave equation
for $\kappa_{ij}$ involving also the background lapse \cite{prd1}. We
then separate the angular dependence in this equation by expanding
$\kappa_{ij}$ in terms of tensor spherical harmonics $(\hat
e_1)_{ij},\ldots,(\hat f_4)_{ij}$ \cite{rw57,m74} (we use the notation
of~\cite{m74})

\begin{eqnarray}
\kappa_{i j} =  a_\times (t,r) (\hat e_1)_{i j} + 
              r b_\times (t,r) (\hat e_2)_{i j} + 
 	     \widetilde N^{-2} a_+ (t,r) (\hat f_2)_{i j} + 
                   r b_+ (t,r) (\hat f_1)_{i j} + 
	\hskip 1.0truecm \nonumber\\
	        r^2 c_+ (t,r) (\hat f_3)_{i j} &+&
		r^2 d_+ (t,r) (\hat f_4)_{i j} \ ,
\label{eq:kappa_expand}
\end{eqnarray}
Note that $(\hat e_1)_{i j},\cdots,(\hat f_4)_{i j}$ are functions
of $(\theta,\phi)$ only and, for clarity, angular
indices $(\ell,m)$ for each mode are suppressed. Similarly, the odd-parity multipoles
$a_\times$ and $b_\times$ and the even-parity multipoles $a_+$, $b_+$,
$c_+$, and $d_+$ also have suppressed indices for each angular
mode.  There is an implicit sum over all angular modes in
(\ref{eq:kappa_expand}).

	The six multipole amplitudes are not independent.
 We use the linearized momentum constraints  to eliminate the
odd-parity amplitude $b_\times$ and the even-parity amplitudes $b_+$
and $c_+$. As a result, for each $(\ell,m)$ mode\footnote{Hereafter we
		will consider only the radiative modes, i.e. those 
		with $\ell \geq 2$}
we  obtain one odd-parity equation for $a_\times$:

\begin{eqnarray}
\Biggl\{ \partial^2_t - \widetilde{N}^4 \partial^2_r - 
	\frac{2}{r}\widetilde{N}^2 \partial_r 
        - \frac{2 M}{r^3} \left(1 - \frac{3 M}{2 r} \right) + 
	\widetilde{N}^2 \left[ \frac{\ell(\ell+1)}{r^2} - 
	\frac{6 M}{r^3} \right] \Biggr\} 
        (a_\times)_{_{\ell m}} = 0 \ ,
\label{oddwave}
\end{eqnarray}
and two coupled even-parity equations for $a_+$ and $h$:
\begin{eqnarray}
	\hskip 0.0truecm \Biggl[ \partial^2_t - \widetilde{N}^4 
	\partial^2_r - \frac{6}{r}\widetilde{N}^4 \partial_r
	+\widetilde{N}^2 \frac{\ell(\ell+1)}{r^2} - \frac{6}{r^2} +
	\frac{14M}{r^3}-\frac{3M^2}{r^4}
	\Biggr] (a_+)_{_{\ell m}} + \hskip 6.0truecm
	\nonumber\\ 
	\Biggl[\frac{4}{r} \widetilde{N}^2 \left(1 -\frac{3M}{r}\right) 
	\partial_r + \frac{2}{r^2} 
	\left(1 - \frac{M}{r} - \frac{3M^2}{r^2}\right) 
	\Biggr] (h)_{_{\ell m}} = 0 \ , 
\label{evenwave1} 
\end{eqnarray}	
\begin{eqnarray}	
	\Biggl[ \partial^2_t - \widetilde{N}^4 \partial^2_r - 
	\frac{2}{r}\widetilde{N}^2 \partial_r
	+ \widetilde{N}^2 \frac{\ell(\ell+1)}{r^2} 
	+ \frac{2 M}{r^3} -
	\frac{7 M^2}{r^4} \Biggr] (h)_{_{\ell m}} 
	- \frac{2 M}{r^3} \left(3 - \frac{7 M}{r}\right) 
	(a_+)_{_{\ell m}} = 0 \ .
\label{evenwave2}
\end{eqnarray}

	The independent multipole amplitudes $(a_\times)_{_{\ell m}}$,
$(a_+)_{_{\ell m}}$, $(h)_{_{\ell m}}$ and the corresponding wave
equations (\ref{oddwave})--(\ref{evenwave2}) for each $(\ell,m)$ mode
are at the basis of the Cauchy-perturbative matching. [Here
$(h)_{_{\ell m}}$ is defined in terms of the trace of $\kappa$,
i.e. $\kappa = (h)_{_{\ell m}}  Y_{_{\ell m}}$ where $Y_{_{\ell
m}}(\theta,\phi)$ is the standard scalar spherical harmonic.]

\subsection{The basic implementation}
\label{tbi}

	As discussed in Section \ref{intro}, with a few
straightforward modifications, this method can be applied to a generic
3D numerical relativity code which solves the Cauchy problem of
Einstein's equations in either the standard $3+1$ or hyperbolic
form. In addition to the standard time integration of the extrinsic
curvature $K_{i j}$ and of the spatial metric $g_{i j}$, three new
procedures are performed during each timestep.

	{\it (A)} The gravitational radiation information contained in
$K_{i j}$ and $\partial_t K_{i j}$ is transformed into the independent
multipole amplitudes $(a_\times)_{_{\ell m}}$, $(a_+)_{_{\ell m}}$,
$(h)_{_{\ell m}}$ and their time derivatives for all of the relevant
$(\ell,m)$ modes (see Paper I for details). The maximum mode at which
the angular decomposition is truncated depends on the basic features
of the problem under investigation. However, a simple comparison of
the relative amplitudes of the different multipoles it is usually
sufficient to provide information about maximum mode necessary.

	{\it (B)} The values of multipole amplitudes and their time
derivatives computed at the extraction 2-sphere on a given timeslice
are imposed as inner boundary conditions on the 1D grid {\tt L} and
evolved [using the radial wave equations
(\ref{oddwave})--(\ref{evenwave2}) for each $(\ell, m)$ mode] forward
to the following timeslice\footnote{Initial
		data on the 1D grid {\tt L} is set consistently with
		the initial data for the 3D grid {\tt N}.} This 
provides the solution, for the new time level, on the whole
perturbative region ${\cal P}$. Since the outer boundary of {\tt L} is
located, by construction, well out in the wave zone, a simple radial
outgoing wave Sommerfeld condition can be imposed there.

	{\it (C)} From the values at the new time level of
$(a_\times)_{_{\ell m}}$, $(a_+)_{_{\ell m}}$, $(h)_{_{\ell m}}$ and
of their time derivatives, it is possible to ``reconstruct'' the
values of $K_{i j}$ or $g_{i j}$ and thus to impose outer boundary
conditions on the 3D grid {\tt N}. The details of how this is done
depend on the formulation of Einstein's equations solved in {\tt
N}. In the computations discussed here, we have used the 3D ``interior
code'' of the Alliance \cite{prl2} adopting an $3+1$ formulation of
Einstein's equations. In this case, only the outer boundary data for
$K_{i j}$ are necessary, since the interior code can calculate $g_{i
j}$ at the outer boundary by integrating in time the boundary values
for $K_{i j}$.

	It is important to  emphasize the great flexibility of the
Cauchy-perturbative approach in providing outer boundary value
data. Once again, the ultimate goal is that of providing, during each
timestep, boundary values of the relevant quantities at the 2-surface
{\bf B} delimiting the 3D grid {\tt N}. Since we can compute the new
values $K_{i j}$ at any point of {\tt N} which lies in the
perturbative region, not only we can provide boundary data on a
2-surface of arbitrary shape, but, (if necessary) on the the whole
portion of {\tt N} outside of the extraction 2-sphere {\bf E}. This
represents a great advantage for which no approximation is required
and, as we will discuss in detail in Section \ref{ltb}, represents an
essential prescription in order to obtain the stability of the code on
very long timescales.

\section{Numerical Setup}
\label{nsu}

	In Paper I we presented tests of our code based on the
propagation of linear waves on a Minkowski background (i.e. with
$M=0$). In those tests, we simulated the Cauchy evolution of the
nonlinear interior code by providing an analytic solution on the 3D
grid. This was necessary in order to evaluate the accuracy and the
convergence properties of the module independently of any error which
may develop due to the numerical evolution. As a result of those
investigations, we were able to show the module's ability to extract
the gravitational wave information, to evolve this information out
to large distances and to impose self-consistent and convergent
Dirichlet outer boundary conditions. While extremely useful, those
tests could not address a number of important questions which are
strictly related to the use of numerical data coming from the solution
of the full Einstein's equations. In particular: (i) what is the
influence of the location of the extraction 2-sphere {\bf E} on the
accuracy of the extracted gravitational wave data?  (ii) do Dirichlet
boundary conditions on {\bf B} provide a long term stability?  (iii)
what are the most convenient boundary conditions to impose?  (iv)
are there numerical techniques that would improve the application of
the Cauchy-perturbative matching? In this paper we provide an answer
to all of these questions and discuss the properties of a
Cauchy-perturbative matching in the more realistic study of linear and
mildly nonlinear waves.

	As in Paper I, we have here computed the propagation of
$\ell=2$, $m=0$ (unless otherwise stated) even-parity linear waves,
initially modulated by a Gaussian envelope with amplitude $A=10^{-6}$
and width parameter $b=1$ \cite{Burke,Teukolsky} (Section \ref{vott}
will discuss variations on this type of initial data by using higher
$\ell$ modes and higher amplitudes). Being time-symmetric at the
initial time, these waves have ingoing and outgoing parts. At each
time level, the extrinsic curvature and 3-metric are computed using
the interior code of the Alliance solving the full Einstein's
equations with a geodesic slicing condition (i.e. $N=1$, $\beta^i=0$)
on a 3D vertex-centered grid, with extents $(x,y,z) \in [-4,4]$. The
code can provide a solution using either an explicit Leapfrog
evolution scheme or a semi-implicit Crank-Nicholson one \cite{nr} and
we will make explicit reference to which of the two we have used in
the different results presented. We have also used a number of
different grid resolutions ranging from $(17)^3$ to $(129)^3$ grid
points and comparable resolutions have been used on the extraction
2-sphere. In the following Sections we will discuss in the detail the
results of our computations and concentrate on two different but
interrelated aspects, namely the ``short term'' and ``long term''
behaviours. In the first we will consider gravitational wave
extraction and imposition of boundary conditions on timescales
comparable with the crossing timescale of the numerical grid (i.e. $t
\sim 8$). In the second section we consider the opposite regime and
investigate the effects of a Cauchy-perturbative matching on time
scales much larger than the crossing timescale (i.e. $t \gg 8$) when
most of the radiation has left the numerical grid and the stability
properties of the module are put to a test.

\section{The short term behaviour}
\label{stb}

	As mentioned in \cite{prl1,prd1}, in the case of a flat
background spacetime (as in the present case) and for weak waves on
Schwarzschild-like backgrounds, the position of the extraction
2-sphere is arbitrary. This gives us the important possibility of
analyzing the influence of the position of the extraction 2-sphere on
the accuracy of the gravitational information read-off, and how this
then affects the accuracy of the boundary conditions which are
provided.

\begin{figure}[h]
\epsfxsize=15.0truecm
\begin{center}
\leavevmode
\epsffile{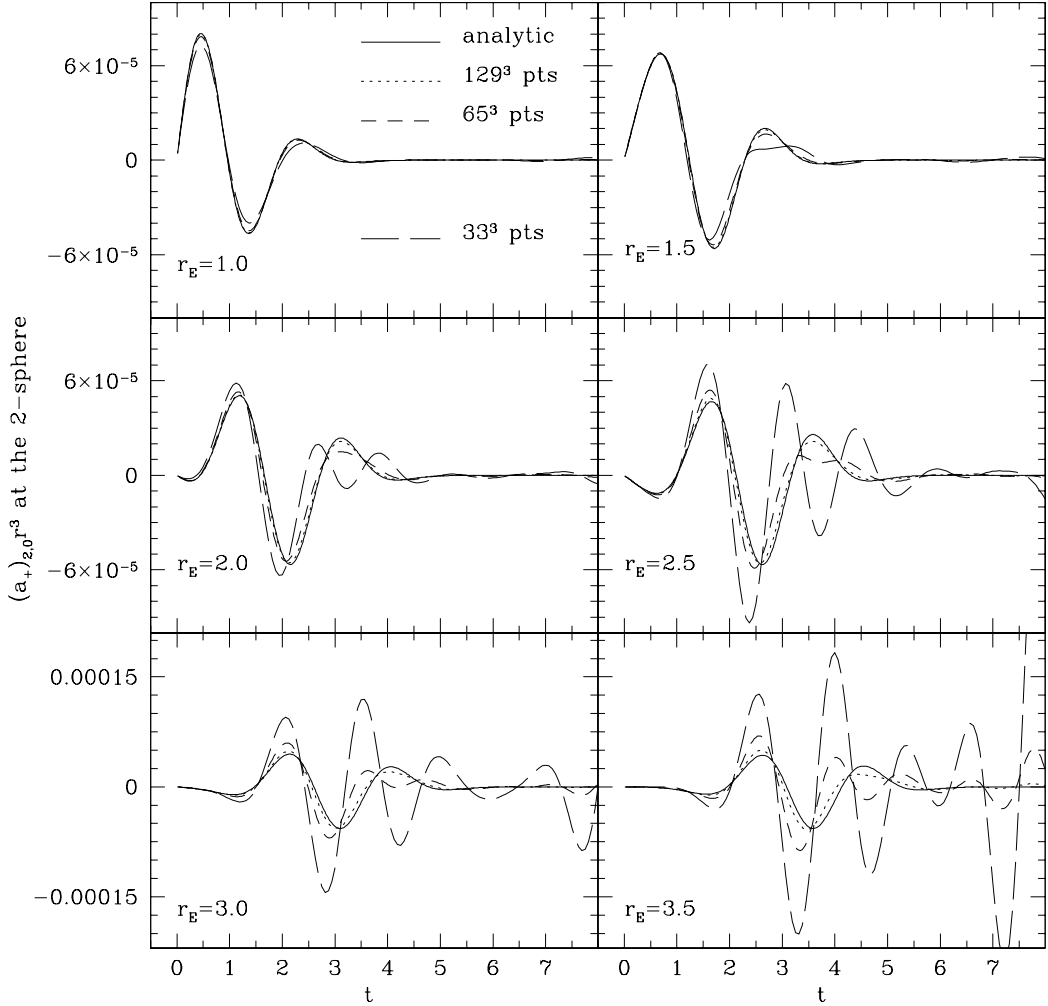}
\end{center}
\caption[fig2]{\label{fig2}
	Timeseries of the multipole amplitude $(a_+)_{_{2 0}}$
	extracted at a 2-sphere of radius $r_{_E} =
	1.0,~1.5,~2.0,~2.5,~3.0,~3.5$. Different grid resolutions are
	indicated with different line types, with a dotted line
	referring to $(129)^3$ grid points, a short dashed line
	referring to $(65)^3$ grid points and a long dashed line
	referring to $(33)^3$ grid points. The analytic solution is
	indicated with a continuous line. Note that we have scaled the
	amplitude by $r^3$ to compensate for the radial fall-off. 
	We have here used a leapfrog integration scheme.}
\end{figure}

	In Fig. \ref{fig2} we show the timeseries of the multipole
amplitude $(a_+)_{_{2 0}}$ extracted at {\bf E} (In the case of an
initial traceless $\ell=2$, $m=0$ wave packet this is the only
analytically non-zero multipole). Other multipoles of the same mode
[e.g. $(a_+)_{_{2 1}}$, $(a_+)_{_{2 -1}}$, $(a_+)_{_{2 2}}$,
$(a_+)_{_{2 -2}}$] as well as other parity amplitudes
[i.e. $(a_\times)_{_{\ell m}}$, $(h)_{_{\ell m}}$] are also extracted,
but their amplitudes are generally several orders of magnitude
smaller. The six different diagrams refer to the six different
positions at which we have placed the extraction 2-sphere
(i.e. $r_{_E} = 1.0, ~1.5, ~2.0, ~2.5, ~3.0, ~3.5$). Each diagram also
shows the same quantity computed at three different resolutions
[namely with $(129)^3$, $(65)^3$ and $(33)^3$ grid points] and we
scale the amplitude by $r^3$ to compensate for the leading-order
radial fall-off of $(a_+)_{_{2 0}}$.

	It is clear from Fig. \ref{fig2} that there is an increasing
relative error between the analytic solution and the extracted data as
the extraction 2-sphere is placed at larger radii while the resolution
is held constant (e.g. compare results at $r_{_E}=1.0$ and
$r_{_E}=3.5$). Since the results shown in Fig. \ref{fig2} do not vary
if the resolution on the two sphere (i.e. the number of grid points
used to cover the extraction 2-sphere) is increased or decreased, the
origin of this behaviour has to be found in the intrinsic numerical
error which is introduced by the solution of Einstein's equation by
the interior 3D code and which becomes larger as the waves propagate
outwards. In fact, a more careful investigation of the behaviour of
the multipole amplitudes other than $(a_+)_{_{2 0}}$ shows that the
initially traceless linear waves develop a non-zero trace of $K_{ij}$
as the evolution proceeds [The presence of non-zero trace becomes
apparent by looking at the amplitudes of the extracted $(h)_{_{\ell
m}}$]. A non-zero trace of $K_{ij}$ is due to truncation error and it
rapidly converges to zero as the resolution is increased, but it has a
subtle effect on the accuracy of the extracted data. While the
multipole amplitudes fall approximately as $\sim r^{-3}$, the non-zero
trace of $K_{ij}$ remains constant during the time evolution. As a
result, for increasing extraction radii, the difference in the
amplitudes of say, $(a_+)_{_{2 0}}$ and $(h)_{_{2 0}}$, becomes
smaller and smaller. For $r_{_E} \gtrsim 3$ the two multipoles are
comparable and this error becomes more severe as a coarser resolution
is used. However, it is also clear that all of these pathologies can
be cured by simply increasing the resolution and Fig. \ref{fig2} shows
the timeseries rapidly converging to the analytic solution as the
resolution is increased even in the most extreme case of an extraction
radius $r_{_E}=3.5$. In view of this, we can summarize the properties
of the {\it perturbative radiation extraction} as follows: for any
extraction 2-sphere location, it is always possible to find a
resolution for the interior grid which will provide gravitational wave
information with the required accuracy.

\begin{figure}[h]
\epsfxsize=15.0truecm
\begin{center}
\leavevmode
\epsffile{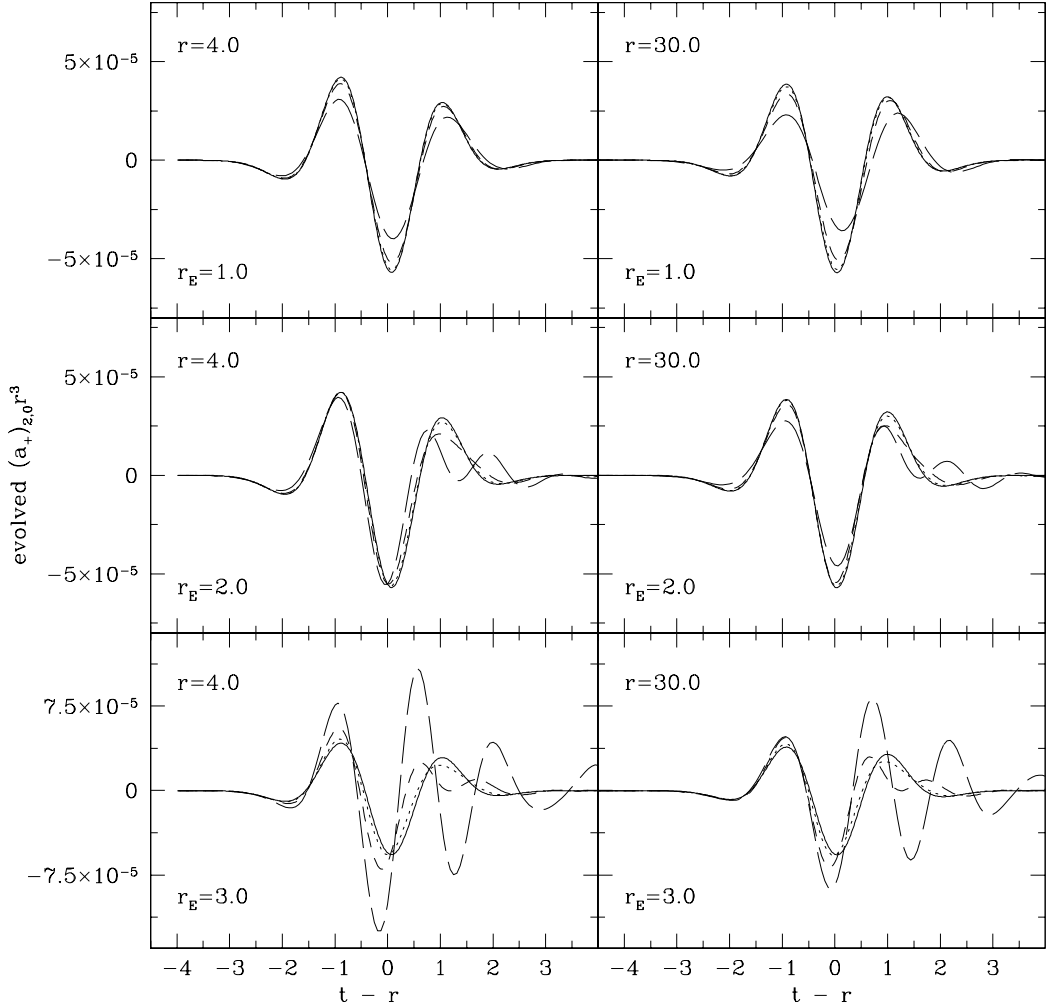}
\end{center}
\caption[fig3]{\label{fig3}
	Timeseries of the multipole amplitude $(a_+)_{_{2 0}}$ evolved
	out to a large distance from the center of the grid.  The
	diagrams on the left show values of $(a_+)_{_{2 0}} r^3$
	evolved out to a radius $r=4.0$ (indicated at the top left
	corner of each diagram), while the diagrams on the right show
	values of $(a_+)_{_{2 0}} r^3$ evolved out to the asymptotic
	radius $r=30.0$. We also show data coming from multipoles
	being extracted at different radii (i.e. $r_{_E} = 1.0, ~2.0,
	~3.0$) and at different grid resolutions. Here also, the
	amplitude is scaled by $r^3$ to compensate for the radial
	fall-off and a leapfrog integration scheme has been used.}
\end{figure}

	Similar considerations apply also for the values of the
multipole amplitude $(a_+)_{_{2 0}}$ which are propagated at large
distances from the center of the grid. Fig. \ref{fig3} shows a set of
timeseries of $(a_+)_{_{2 0}}$ ``evolved'' out at a radius $r=4.0$
(indicated at the top left corner of each diagram) corresponding
roughly with the outer edge of the 3D grid {\tt N} and at an
asymptotic radius $r=30.0$. Also in this case, we show data extracted
for different positions of the 2-sphere (i.e. $r_{_E} = 1.0, ~2.0,
~3.0$) and at different grid resolutions. Note that the waveforms
evolved at $r=4.0$ and those at $r=30.0$ do not differ significantly
 because at these radii the waveform is dominated by its  
asymptotic part.

	Having shown the the ability of the module to extract
convergent gravitational wave data from a fully nonlinear 3D numerical
relativity code, we next turn to examine the corresponding ability to
``reconstruct'' the extrinsic curvature $K_{ij}$ from the information
extracted at the 2-sphere (cf. Fig. \ref{fig2}) and evolved out to the
asymptotic region (cf. Fig. \ref{fig3}). For the computations
discussed in this Section, we have chosen to impose the simplest type
of boundary conditions that can be implemented within a
Cauchy-perturbative approach. From the solution at the new time level
of the evolution equations (\ref{oddwave})--(\ref{evenwave2}) for each
$(\ell,m)$ mode we calculate the value of the extrinsic curvature at
all the grid points on {\bf B}. We then impose these values as the
outer boundary conditions and will refer to this implementation as the
{\it Dirichlet injection} to distinguish it from other types of
boundary conditions which will be discussed in the next section.

\begin{figure}[h]
\epsfxsize=15.0truecm
\begin{center}
\leavevmode
\epsffile{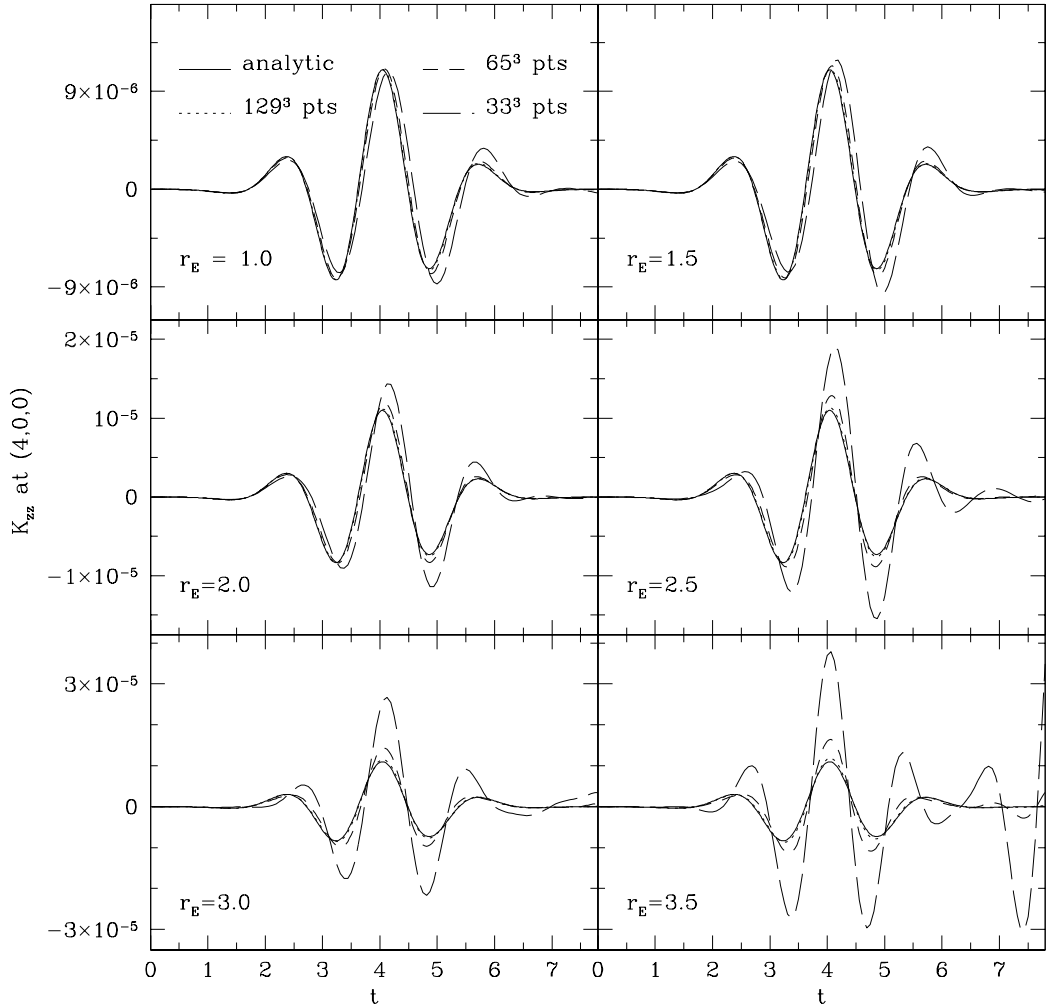}
\end{center}
\caption[fig4]{\label{fig4}
	Timeseries of the reconstructed values of $K_{zz}$. The six
	different diagrams refer to the six different positions of the
	extraction 2-sphere, and for each diagram, results obtained
	with  different grid resolutions are indicated with
	different line types. The timeseries is computed on a
	gridpoint at the outer boundary aligned on the $x-$axis
	[i.e. at a coordinate location $(4,0,0)$] and a leapfrog 
	evolution scheme has been used.}
\end{figure}

\begin{figure}[h]
\epsfxsize=15.0truecm
\begin{center}
\leavevmode
\epsffile{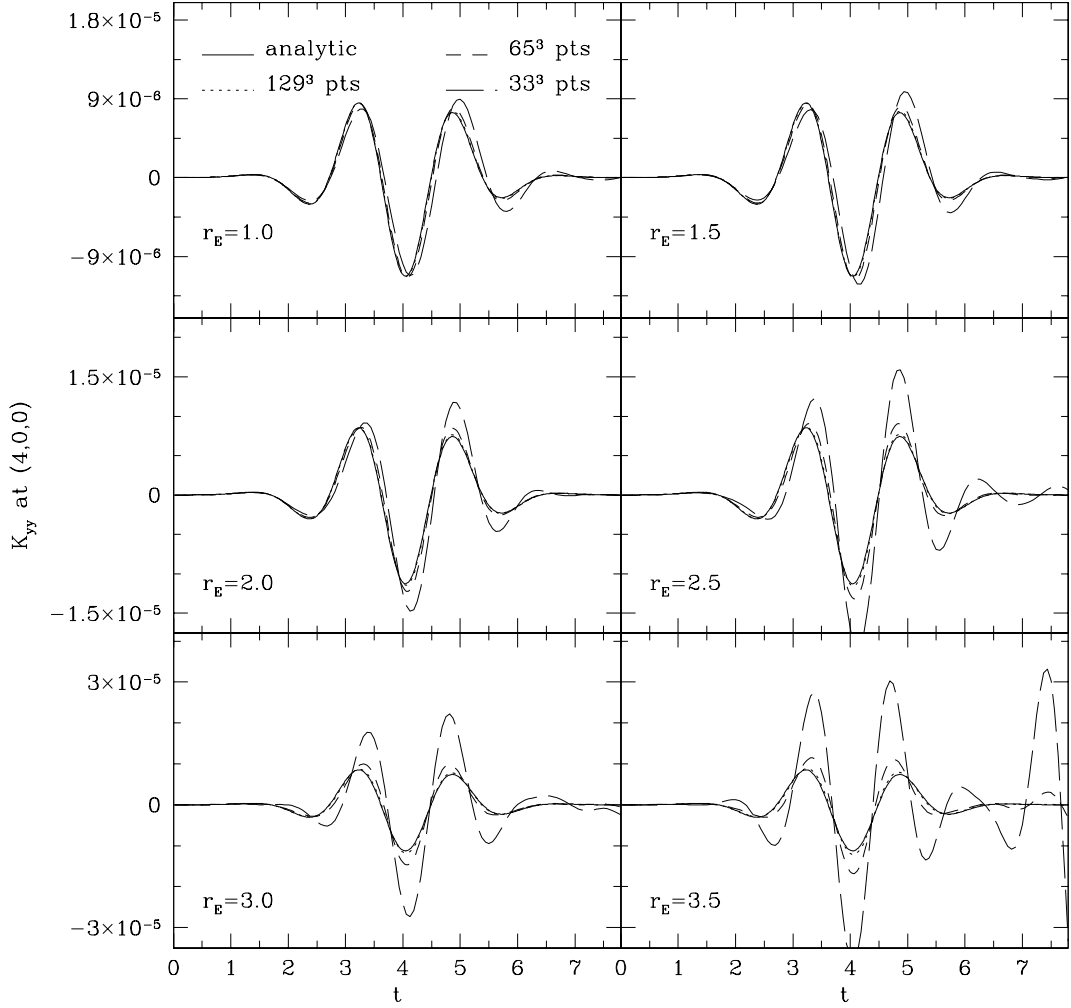}
\end{center}
\caption[fig5]{\label{fig5}
	As in Fig. \ref{fig4} but for the component $K_{yy}$ of the
	extrinsic curvature.}
\end{figure}

	Fig. \ref{fig4} shows a timeseries of the reconstructed values
of the $K_{zz}$ component of the extrinsic curvature
for different positions of the extraction 2-sphere and different grid
resolutions. The timeseries is computed on a gridpoint at the outer
boundary aligned on the $x-$axis [i.e. at a coordinate location
$(4,0,0)$]. Equivalent timeseries are shown in Fig. \ref{fig5} for the
$K_{yy}$ component of the extrinsic curvature, which is basically in
phase opposition with $K_{zz}$.

	It is not surprising that the behaviour of the timeseries for
$K_{ij}$ in both Fig. \ref{fig4} and Fig. \ref{fig5} mimics the one
seen in Fig. \ref{fig2} and Fig. \ref{fig3} for the extracted
multipole amplitudes. In fact, as we will discuss more in the
following section, the extraction 2-sphere and the outer boundary are
closely coupled. The accuracy of the boundary conditions imposed is
clearly dependent on the accuracy of the extracted gravitational wave
information. A large relative error between the extracted and analytic
data will translate into a proportionally large relative error between
the injected values for $K_{ij}$ and the analytic values for the same
quantities.

	It is also important to emphasize that the imposition of poor
boundary conditions does, in turn, produce spurious reflection of
radiation at the outer boundary. This reflected gravitational wave
information will contaminate the radiation signal read off at the
extraction 2-sphere leading to increasingly larger differences from
the purely outgoing analytic solution.  In a loose sense, the
extraction 2-sphere and the outer boundary behave as a coupled system
of microphones and loud-speakers with the 2-sphere playing the role of
the microphones.  It is clear that such a coupling can be extremely
delicate and might be the cause for exponentially growing
instabilities as we will discuss in the following Section, where we
also indicate a number of prescriptions that make this coupling less
important.

\begin{figure}[h]
\epsfxsize=10.0truecm
\begin{center}
\leavevmode
\epsffile{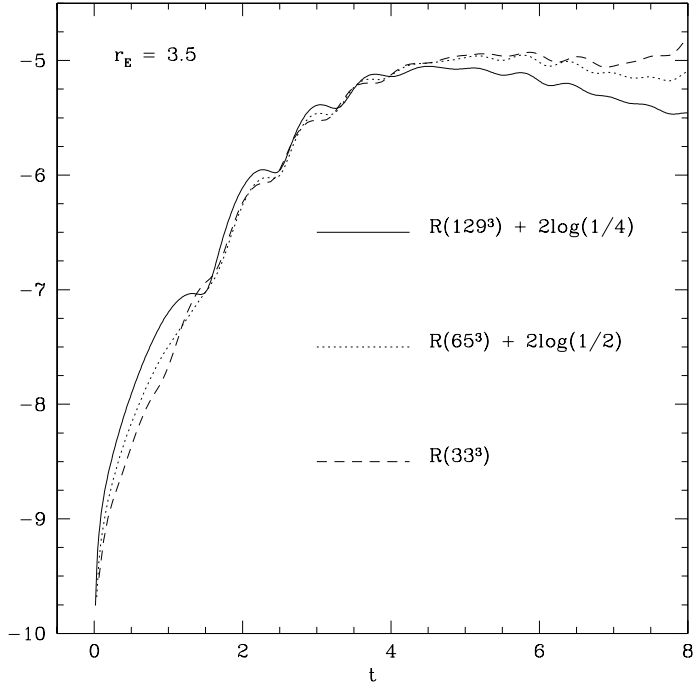}
\end{center}
\caption[fig6]{\label{fig6}
	Logarithm of the $L_2$ norm of the error in $K_{z z}$ computed
	over the whole outer boundary {\bf B} for successive grid
	resolutions and for the most unfavorable position of the
	extraction 2-sphere (i.e. $r_{_E}=3.5$). The differences
	between norms at different grid resolutions are normalized by
	factors $2 {\rm log} (h_{i+1}/h_i)$, where $h_{i+1}, h_{i}$
	are two successive grid resolutions. The overlap of the curves
	shows the second-order convergence of the module. The 
	evolution scheme used is leapfrog.}
\end{figure}

	In Fig. \ref{fig6} we show a more global measure of the
accuracy and of the convergence properties for the boundary data by
computing the $L_2$ norm of the error in $K_{i j}$ as measured over
the whole 3D outer boundary {\bf B}. In particular, in Fig.
\ref{fig6} we plot the $L_2$ norm of the error in $K_{zz}$ at the
outer boundary and for successive grid resolutions (The norms are
scaled logarithmically.). Moreover, in order to make the errors
comparable, we scale the different curves by numerical factors of the
form $2 {\rm log} (h_{i+1}/h_i)$, where $h_{i+1}, h_{i}$ are two
successive grid resolutions [$(h_{i+1}/h_i) = 1/2$ in these tests]. In
order to make this a much more stringent test, we have chosen an
extraction radius $r_{_E}=3.5$. The overlap of the curves is a clear
signature of the second-order convergence of the module and of the
interior code even when the numerical errors are most severe.

\section{The long term behaviour}
\label{ltb}

	Providing an accurate and numerically convergent approximation
to the gravitational waveform in the wave zone surrounding an isolated
source represents a very important feature of any radiation-extraction
and outer-boundary module. However, even more important is that the
module provides a stable evolution of the interior equations,
minimizing the numerical reflection of radiation at the boundary.
Particularly for systems of evolution equations in which radiation and
background dominated metric/extrinsic curvature variables are not
easily defined, matching techniques may be the only way to achieve a
stable evolution. In this section we discuss a number of approaches
which were applied to this problem in the context of the Alliance
interior evolution code.

\subsection{Perturbative Sommerfeld boundary conditions}
\label{ps}

	The results presented in the previous section were obtained by
imposing as outer boundary values for $K_{ij}$, the ones reconstructed
from the values, at the new time level, of the multipole amplitudes.
Although straightforward to implement and very accurate, a ``Dirichlet
injection'' of outer boundary data leads to a rapid error growth when
the time evolution is carried for sufficiently long periods of
time. A careful analysis has revealed that these boundary conditions
seem to produce a rather large amount reflection as the gravitational
waves leave the numerical grid. This is basically the result of the
slight mismatch between the wave phases and amplitudes imposed at the
outer boundary and those in the interior of the 3D numerical grid.

	It is clear that a finite discretization will always produce a
certain amount of reflection. It is thus important to study and
develop new techniques that tend to suppress this reflection and
allow, as much as possible, the outgoing radiation to escape freely to
infinity. Interesting results in this direction have been obtained
implementing a boundary condition we have named {\it perturbative
Sommerfeld} \cite{prl1}. While a simple Sommerfeld outgoing wave
condition applied to a tensor quantity such as the extrinsic curvature
or the three metric is (strictly speaking ) incorrect, it is often
satisfactory, especially if applied in the distant wave zone (cf.
\cite{zerilli98,sc97} for the case of linear waves). In view of this,
we have related, at the outer boundary {\bf B}, the null derivatives
of the extrinsic curvature obtained from the interior grid
(i.e. $K_{ij}$) with the one obtained from the perturbative module
($\kappa_{ij}$, since background extrinsic curvature is assumed to be
zero):

\begin{equation}
\label{pse}
{\partial \over \partial t} \left(K_{ij} - \kappa_{ij} \right) +
{\partial \over \partial r} \left(K_{ij} - \kappa_{ij} \right) +
{q \over r} \left(K_{ij} - \kappa_{ij} \right) = 0 \ ,
\end{equation}
where $q$ is a positive integer\footnote{
		Results do not depend sensitively on the value of $q$,
		which we have chosen to be 2 in these tests in order 
		to reproduce the leading order term in the asymptotic 
		radial fall-off.}. 
In other words, we ``correct'' the Sommerfeld outgoing wave condition
${\partial_t} K_{ij} + \partial_r K_{ij} + (2/r) K_{ij} = 0$, with a
right-hand side which is usually taken to be zero but which is not (in
general) zero. As a result, this prescription resembles a Sommerfeld
condition but is effectively much more general since (i) it can be
used in regions where the radiation is not dominated by the asymptotic
outgoing behavior and (ii) it takes into account arbitrary angular
dependence, as well as the effects of a Schwarzschild black hole
background. Since the perturbative correction can be very small and is
the result of a near cancelation of several terms involving space and
time derivatives, it is important to implement equation (7) so that
the same numerical differential operator acts on both $K_{ij}$ and
$\kappa_{ij}$.
	
\begin{figure}[h]
\epsfxsize=10.0truecm
\begin{center}
\leavevmode
\epsffile{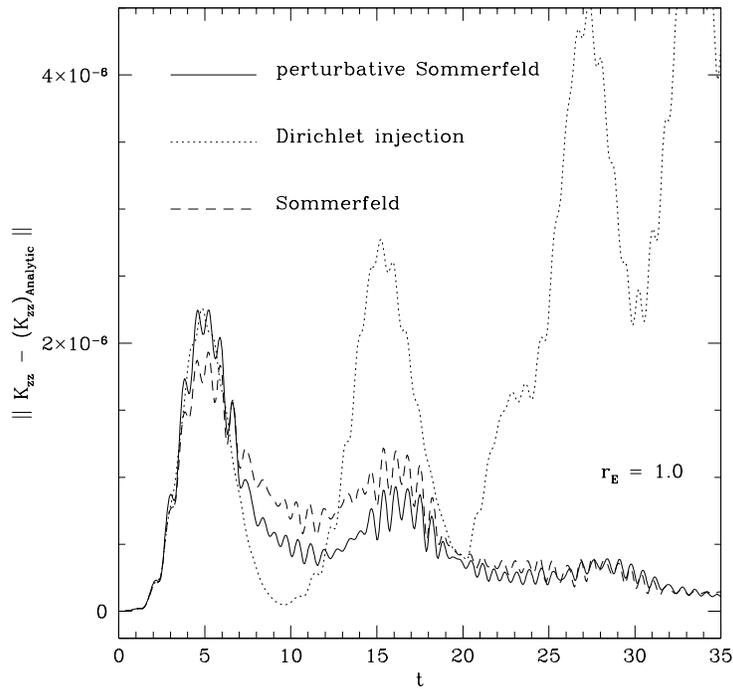}
\end{center}
\caption[fig7]{\label{fig7}
	Timeseries of the $L_2$ norm of the error in $K_{zz}$ for an
	extraction at $r_{_E}=1.0$. The interior code has a resolution
	of $(33)^3$ grid points and uses a leapfrog integration
	scheme. The different curves refer to the
	different types of boundary conditions used. }
\end{figure}

	Implementation of this method has shown that the perturbative
Sommerfeld approach is very accurate and generally yields longer
evolutions than the direct injection of Dirichlet data. Figure
\ref{fig7} shows a direct comparison of three different boundary
conditions, namely, perturbative Sommerfeld, Dirichlet injection and
Sommerfeld. In particular, we show the timeseries of the $L_2$ norm of
the error in $K_{zz}$ for an extraction at $r_{_E}=1.0$. Because we
were interested in results over very long times we were forced to
perform computations using a very coarse resolution of $(33)^3$ grid
points.

	There are a number of interesting features that emerge from
Fig. \ref{fig7}. The most evident one is strikingly different
behaviour between a direct injection of reconstructed data and the use
of a Sommerfeld-like boundary condition. In the case of a Dirichlet
injection, in fact, not all of the radiation is able to leave the
numerical grid, but some of it remains trapped and is repeatedly
read-off. In this case the coupling between the extraction 2-sphere
and the outer boundary is very strong and amplifies the numerical
error which grows exponentially in time, with a beat frequency roughly
set by the dimensions of the numerical grid. The perturbative
Sommerfeld and the Sommerfeld conditions, on the other hand, are much
more effective in letting the radiation escape off the mesh and
whatever the amount of reflection, this is progressively damped as the
evolution proceeds. In this respect, a perturbative Sommerfeld
condition is more efficient in suppressing the reflected incoming
radiation (e.g. for $7 \lesssim t \lesssim 17$) and seems to behave
basically like a Sommerfeld condition as the evolution progresses
further to intermediate times.

\begin{figure}[h]
\epsfxsize=10.0truecm
\begin{center}
\leavevmode
\epsffile{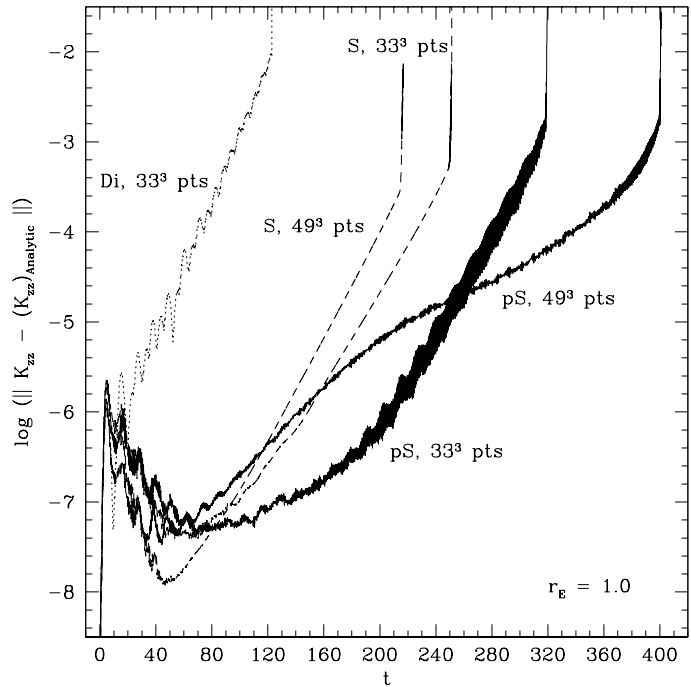}
\end{center}
\caption[fig8]{\label{fig8}
	As in Fig. \ref{fig7} but for longer timescales and the use of
	the logarithms of the norms. The different
	curves refer respectively to Dirichlet injection boundary
	conditions (Di), Perturbative Sommerfeld boundary conditions
	(pS) and Sommerfeld boundary condition (S). Curves with a
	higher resolution of $(49)^3$ grid points are also plotted for
	the perturbative Sommerfeld and the Sommerfeld
	boundary conditions. All curves have been obtained with a
	leapfrog integration scheme.}
\end{figure}

	However, the intrinsically different character of the
perturbative Sommerfeld condition becomes apparent on much longer
timescales or with higher resolutions. Fig. \ref{fig8} is the same as
Fig. \ref{fig7} but on a longer timescale. Also, additional curves
computed with a resolution of $(49)^3$ grid points are shown for a
comparison between the perturbative Sommerfeld and the Sommerfeld
condition. While the two types of boundary conditions do not seem to
differ significantly and both show the emergence of exponentially
growing errors, the use of perturbative boundary conditions delays the
onset of the error growth and allows for a much longer evolution.
Moreover, by increasing the interior resolution we can further prolong
the running time. This is in stark contrast to the behaviour of the
Sommerfeld condition, for which an increased interior resolution
results in a shorter running time \cite{prl1}. As shown in
Fig. \ref{fig8}, by using $(49)^3$ interior gridpoints, we were able
to evolve the code up to $t \sim 400$ and about four times longer than
for the case of Dirichlet injection. This is a comparatively very
long timescale, which is more than $50$ times longer than the
physically relevant one, i.e. the crossing timescale.

	At present, it is not clear what is the origin of the
exponential error growth observed in Fig. \ref{fig7} and which appears
also with evolutions using a harmonic slicing of spacetime. Such
instabilities might be directly related to a nonlinear coupling
between waves reflected off the outer boundary and numerical
instabilities triggered by the $3+1$ form of Einstein's equations. It
is indeed remarkable that no exponential growth is present in other
formulations of Einstein's equations, such as the one proposed by
Shibata and Nakamura or the Einstein-Ricci hyperbolic formulation, in
which linear waves have been stably evolved in harmonically sliced
spacetimes \cite{sn95,cs97}.

\subsection{Perturbative boundary conditions with ``blending'' }
\label{pb}

	The perturbative Sommerfeld boundary conditions represent a
very promising implementation of the Cauchy-perturbative matching and
provide both high accuracy and very small reflection at the outer
boundary. Despite these appealing features, they do not provide for
very long term stability of the interior code and, as shown in Fig
\ref{fig8}, they eventually suffer from an exponential growth. Of
course, in the computations reported here, the boundary conditions
provided by the Cauchy-perturbative matching are totally adequate on
the timescale necessary for the gravitational waves to leave the
numerical grid. Indeed, the instabilities induced by the outer
boundary become relevant only long after the crossing timescale, when
the grid basically contains numerical noise. However, providing
accurate boundary conditions on a dynamical timescale is usually not
sufficient and the achievement of unconditionally stable codes is not
only of academic interest. In order to successfully model the problem
of binary black hole coalescence, the numerical code will have to be
able to stably solve Einstein's equations on a timescale (the
gravitational wave emission one) which is much longer than the
dynamical one. It is therefore imperative to devise a technique which
provides unconditionally stable boundary conditions. A first
successful step in this direction has been made with the
implementation of the ``perturbative-blending'' technique we will
discuss in this Section.

\begin{figure}[h]
\epsfxsize=10truecm
\vskip 1.0truecm 
\begin{center}
\leavevmode
\epsffile{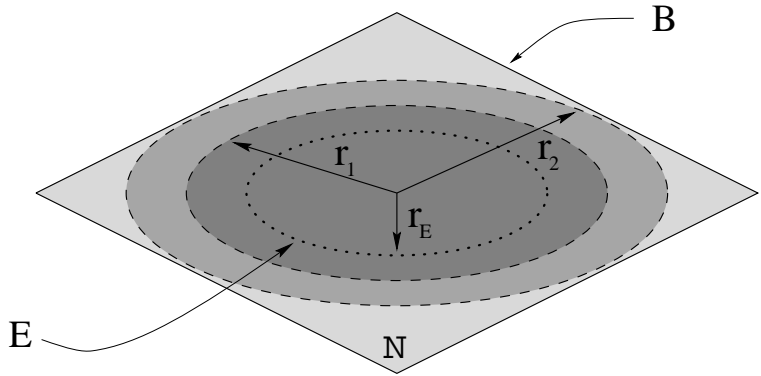}
\end{center}
\vskip 1.0truecm 
\caption[fig9]{\label{fig9}
	Schematic of the numerical implementation of a ``perturbative
	blending'' matching. As in Fig. \ref{fig1}, {\tt N} represents
	the 3D numerical grid in which the full Einstein's equations
	are solved and {\bf B} its 2D outer boundary. $r_{_E}$ is the
	radius of the extraction 2-sphere {\bf E} and $r_1, r_2$ are
	two radii chosen so that $r_2 > r_1 > r_{_E}$. The dark shaded
	region (i.e. $r < r_1$) shows the portion of {\tt N} filled by the
	numerical nonlinear solution provided by the interior
	code. The medium shaded region (i.e. $r_2 < r < r_1$) shows
	the portion of {\tt N} in which the solution of the interior
	code is ``blended'' with the one coming from the linearized
	Einstein's equations. Finally, in the light shaded region,
	only the perturbative solution is used.}
\end{figure}

	As discussed in the previous Section, the perturbative
Sommerfeld boundary conditions are much more effective in providing
accuracy and reducing reflection at the outer boundary than is the
simple Dirichlet injection. One of the major differences between the
two types of boundary conditions is in their numerical implementation.
This involves only the outermost boundary grid points (i.e. those on
{\bf B} of Fig. \ref{fig9}) in the case of a Dirichlet injection but
also the closest interior neighbouring gridpoints in the case of a
perturbative Sommerfeld condition [necessary for taking a finite
difference form of the spatial derivatives in (\ref{pse})]. The
perturbative blending can then be considered as the extension of the
perturbative Sommerfeld condition to a larger set of gridpoints. The
basic idea is simple and based on the attempt of modifying the
propagation characteristics in the vicinity of the outer boundary with
the goal of acting distinctively on the outgoing and ingoing parts of
the gravitational waves. A detailed description of the basic
properties of {\it ``sponge filters''} in conjunction with absorbing
boundary conditions in one-dimensional wave propagation can be found
in \cite{io81} and results of its application are also discussed in
\cite{mc96}. The interpretation of the blending as an implementation
of the sponge filter method will be given in Appendix A.

	Figure \ref{fig9} gives a schematic representation of the way
the perturbative blending has been implemented. The key feature of
this specific approach comes from exploiting the module's ability to
provide a perturbative solution to Einstein's equation not only on the
outer boundary {\bf B} but, in principle, in the whole region of the
3D numerical grid {\tt N} outside of the extraction 2-sphere {\bf
E}. However, rather than doing this, we have isolated a spherical
shell of radii $r_1$ and $r_2$ (where $r_2 > r_1 > r_{_E}$) and
blended therein the nonlinear solution coming from the interior code
with the linear one coming from the Cauchy-perturbative module (this
is shown as the medium shaded region in Fig. \ref{fig9}). In
particular, at the end of each time step (as well as during each
iteration of the Crank-Nicholson evolution scheme we have used in
these tests), we do the following: at all of the gridpoints at $r <
r_1$ (dark shaded region in Fig. \ref{fig9}) the nonlinear solution is
left unmodified; for all of the gridpoints between $r_1$ and $r_2$ we
``blend'' the nonlinear solution with the perturbative one
reconstructed at that grid point; for all of the gridpoints at $r >
r_2$ we replace the computed values of $K_{i j}$ with perturbative
data.  The ``blending'', consists of smoothly weighting the nonlinear
and linear solutions so that the first one is imposed at the 2-sphere
of radius $r_1$ and the second one is imposed at $r_2$ (see 
Appendix A for details).\footnote{The idea of blending boundary
		conditions was first proposed within
		the Alliance by R. G\'omez and the Pittsburgh group 
		who obtained stable evolution of linear waves after
		blending the numerical solution with the {\it analytic}
		one \cite{rg97}.} 
Stability does not depend on the form of the weighting power function
as long as the latter satisfies the boundary conditions of being zero
at the inner blending shell and one at the outer blending
shell. However, a more careful matching of the first and second
derivatives at the two blending shells does provide a smaller amount
of reflection off the outer boundary during the initial stages of the
evolution (i.e. for $t \lesssim 20$).

\begin{figure}
\epsfxsize=14.5truecm \epsfysize=9.0truecm
\begin{center}
\leavevmode
\epsffile{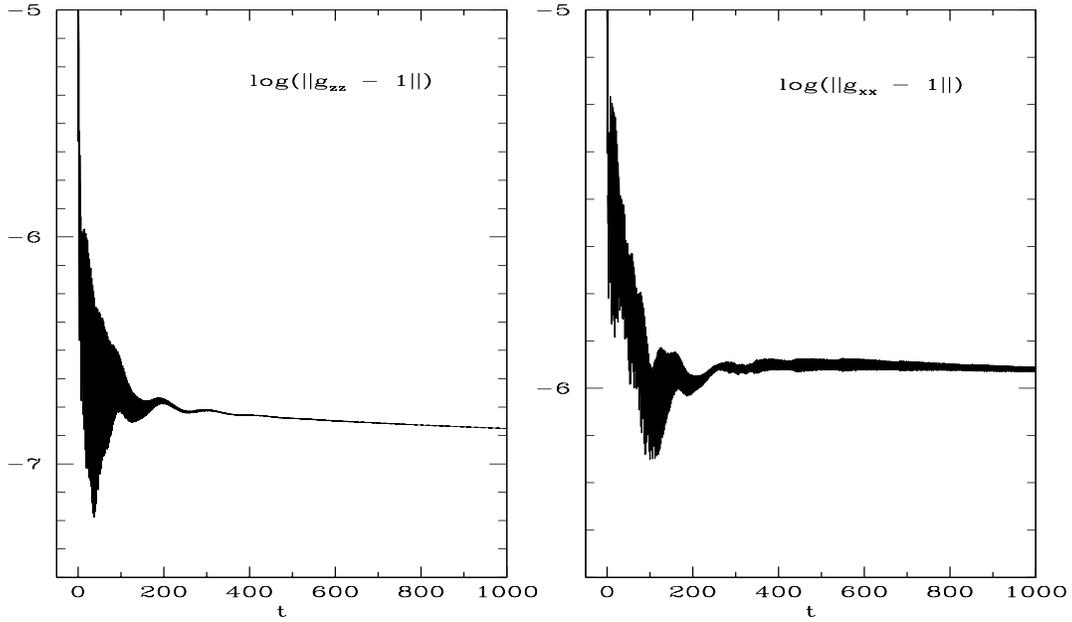}
\end{center}
\caption[fig10]{\label{fig10}
	Timeseries of the errors for the $g_{zz}$ and the $g_{xx}$
	components of the three-metric with the use of a perturbative
	blending match. The norms shown here are taken along the
	$z-$axis and not on a 2D surface as for the previous diagrams.
	In these runs we have used a very coarse resolution of
	$(33)^3$ gridpoints and an extraction radius $r_{_E}=1$. The
	blending region is covered with 10 gridpoints, but similar
	results have been obtained also with smaller numbers of
	gridpoints. The integration scheme used is Crank-Nicholson.}
\end{figure}

	Figs. \ref{fig10} and \ref{fig11} illustrate the radical
changes in the long time behaviour introduced by the use of a
perturbative blending match.

	In particular, in Fig. \ref{fig10} we show the timeseries of
the errors for the $g_{zz}$ and the $g_{xx}$ components of the
three-metric, while in Fig. \ref{fig11}, we show the timeseries of the
errors in the $K_{zz}$ component of the extrinsic curvature and of the
violation of the Hamiltonian constraint. The plots refer to
computations performed using a coarse resolution of $(33)^3$
gridpoints and an extraction radius $r_{_E}=1$.

\vbox{
\begin{figure}
\epsfxsize=14.5truecm
\epsfysize=9.0truecm
\begin{center}
\leavevmode
\epsffile{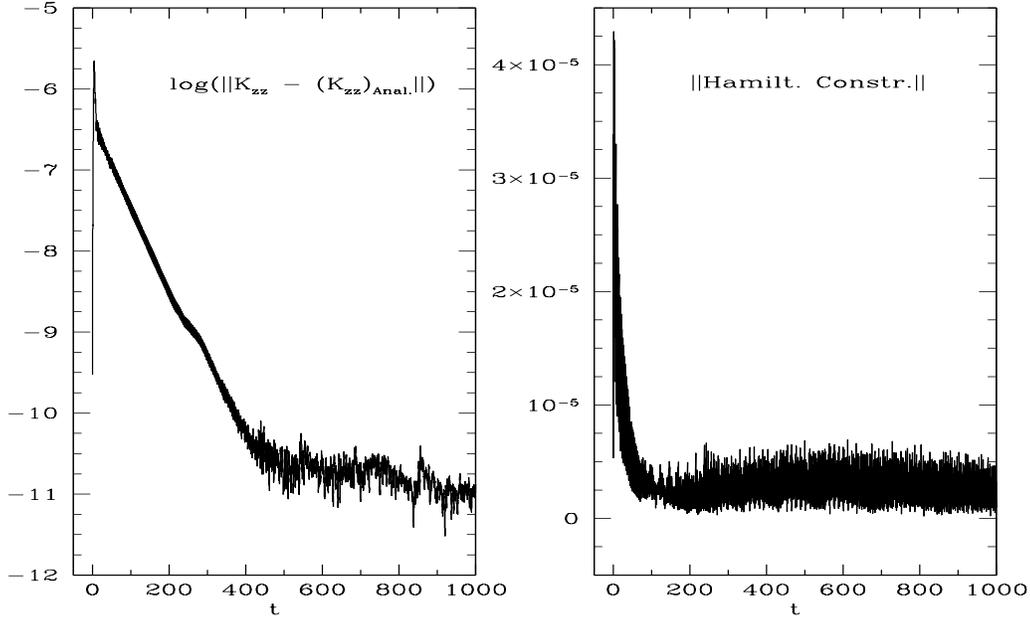}
\end{center}
\caption[fig11]{\label{fig11}
	Timeseries of the errors in the $K_{zz}$ component of the
	extrinsic curvature and of the violation of the Hamiltonian
	constraint when a perturbative blending match is used. The
	resolution is that of $(33)^3$ gridpoints and the extraction
	radius if $r_{_E}=1$ and the blending region is covered with
	10 gridpoints. The integration scheme used is Crank-Nicholson.}
\end{figure} }

	The long evolution times reached and the quiescent behaviour
of the evolved variables reduced to their roundoff error values,
clearly show that the use of a perturbative blending match does
provide the desired long term stability. This is evident from the
gradual decay of the norms and the stationary behaviour of the
violation of the Hamiltonian constraint which does not show any sign
of instability on a timescale more than 125 times the crossing
timescale \cite{lr98}. We have also verified that the
perturbative-blending boundary conditions provide \textit{accurate}
short term extraction of the waveform, comparable to the results in
Figures \ref{fig2}--\ref{fig5}.

	It is also clear that the use of a perturbative blending
introduces two new ``free'' parameters (i.e. the radii $r_1$ and
$r_2$) and a satisfactory implementation will therefore depend on some
``tuning'' and experimentation. In particular, for the runs shown
above we have chosen $r_{_E}=1$, $r_1=2$ and $r_2=4$, which is the
radius of the sphere inscribed in {\tt N} and tangent to it. This
gives about 10 gridpoints along the axes where the blending between
the nonlinear and linear solutions is made. Very similar results have
been obtained also with 9, 8 and 7 gridpoints, but a blending over 6
or fewer gridpoints would make exponentially growing instabilities
reappear. Provided that the intrinsic length of the blending is region
is kept constant, stability has been obtained also with simulations
using a larger or a coarser resolution than the one shown for in
Figs. \ref{fig10} and \ref{fig11}.

	A detailed understanding of the properties of the perturbative
blending matching is still under development and is particularly hard
to achieve given the three-dimensionality of the full
problem. However, there are some basic features that seem to be well
established and that we have illustrated in the Appendix. There, using
simplified 1D model describing the evolution of linear waves on a flat
background, we show that imposing boundary conditions using a mixture
of a numerical solution of Einstein's equations with another one
(either analytic or obtained from a perturbative matching) is
equivalent to imposing a variable phase velocity in the zone where the
blending is made. If appropriate boundary conditions are applied to
this variable phase velocity, the blending can tilt the ingoing
characteristic toward the advected one and prevent ingoing modes
propagating from the outer boundary. Moreover, the blending
progressively dampens the propagation of outgoing modes which are
totally absorbed at the outer boundary. In this way, it is possible to
decouple the outer boundary from the interior evolution without having
to place it at very large distances (see the Appendix for details).

	The use of perturbative blending boundary conditions has
provided the unconditional long-term stability we were requiring to
our radiation-extraction and outer-boundary module. Given the
versatility of the perturbative matching, this approach could
represent a very powerful tool also in other numerical relativity
applications. Further work is necessary in this direction and
experimentation with more complex physical configurations. It is
interesting to note that G\'omez's original prescription of blending
numerical and analytic data has found a partially successful
application also in the 3D Cauchy evolution of a single black hole
where it lead to the first stable evolution \cite{mh98} of this type.

\section{Variations on the theme: high mode and mildly nonlinear waves }
\label{vott}

	In this concluding Section we provide further evidence
of the robustness and versatility of the Cauchy-perturbative matching
in extracting gravitational wave information and providing outer
boundary conditions. For this purpose we present results
obtained from computations having different initial data than the one
discussed so far and concentrate on short term evolutions. 

\begin{figure}[h]
\epsfxsize=12.0cm
\epsfysize=10.0cm
\begin{center}
\leavevmode
\epsffile{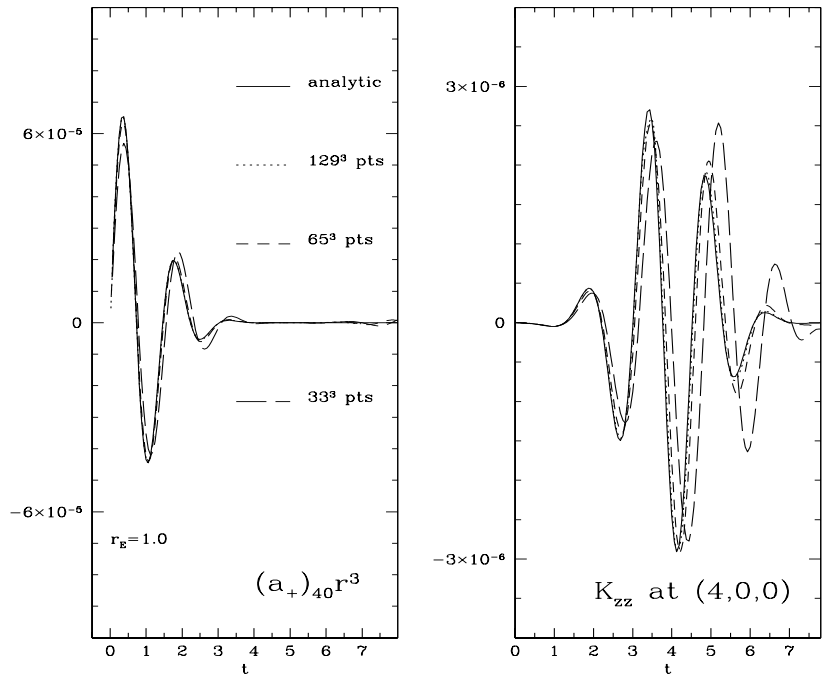}
\end{center}
\caption[fig12]{\label{fig12}
	The left diagram shows the timeseries of the multipole
	amplitude $(a_+)_{_{4 0}}$ extracted at a 2-sphere of radius
	$r_{_E} = 1.0$ for different grid resolutions
	(cf. Fig. \ref{fig2}). The right diagram, the timeseries of
	the reconstructed values of $K_{zz}$ as measured at the
	gridpoint (4,0,0) (cf. Fig. \ref{fig4}).}
\end{figure}

	In particular, in Fig. \ref{fig12} we show the extracted
signal (left diagram) and the injected boundary conditions for
$K_{zz}$ (left diagram) when an initial $\ell=4$, $m=2$ even-parity
wave is used as initial data (We have here maintained the same
amplitude $A=10^{-6}$ and width parameter $b=1$; note also that we
have used the Crank-Nicholson evolution scheme.) Fig. \ref{fig12}
should be compared with Figs. \ref{fig2} and \ref{fig4} where similar
data is reported in the case of an initial $\ell=2$, $m=0$ packet. It
seems evident that also in the case of this higher mode initial data
the Cauchy-perturbative module is able to provide convergent wave
extraction and boundary conditions (For this test and the following
ones in this Section, we have used the computationally less expensive
perturbative Sommerfeld boundary conditions.). 

	Next, we can consider the behaviour of the module when the
initial amplitude of the wave packet is increased. Of course, the
analytic form for the initial data used in these tests is derived in
the linearized regime and a wave packet with an exceedingly large
amplitude will no longer satisfy the Hamiltonian and momentum
constraints. However, we can progressively increase the initial
amplitude and exclude amplitudes above which the violation of the
constraints becomes too severe (e.g. more than 50\%). This allows us
to perform an interesting check of the efficiency of the module in the
linear and mildly nonlinear regime.

\begin{figure}[h]
\epsfxsize=12.5cm
\begin{center}
\leavevmode
\epsffile{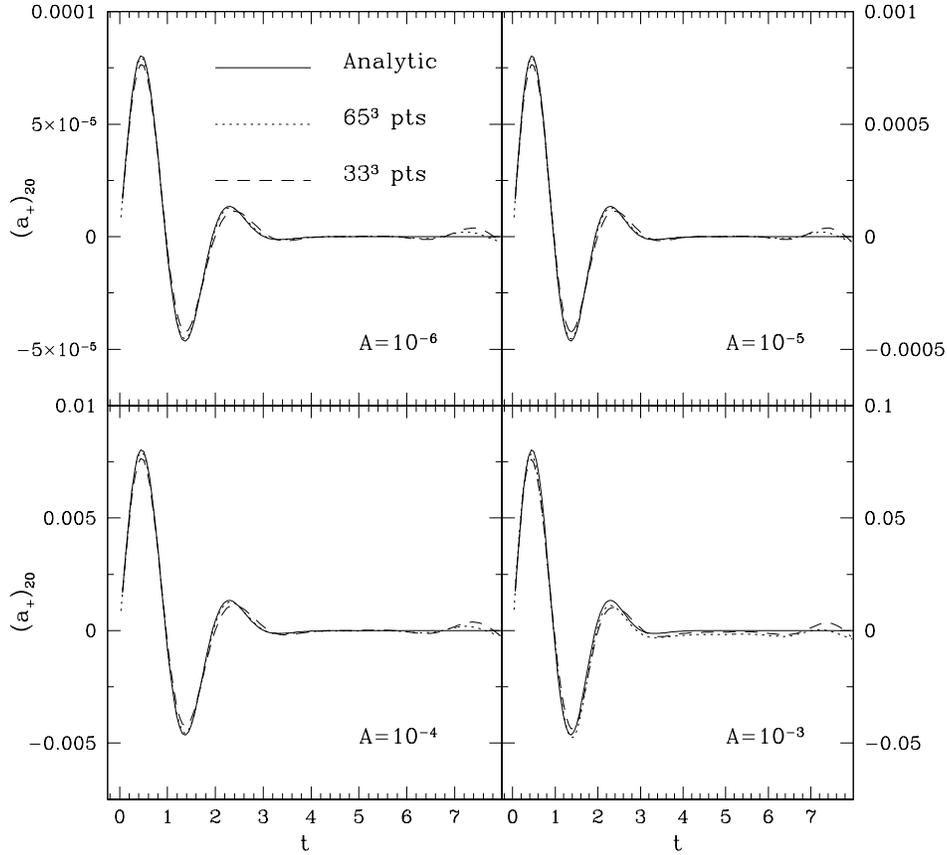}
\end{center}
\caption[fig13]{\label{fig13}
	Timeseries of the multipole amplitude $(a_+)_{_{2 0}}$
	extracted at a 2-sphere of radius $r_{_E} = 1.0$ for different
	values of the initial wave amplitude (i.e. $A =
	10^{-6}-10^{-3}$). Different grid resolutions are indicated
	with different line types (cf. Fig. \ref{fig2}). The
	integration scheme used is Crank-Nicholson.}
\end{figure}

\begin{figure}[h]
\epsfxsize=12.5cm
\begin{center}
\leavevmode
\epsffile{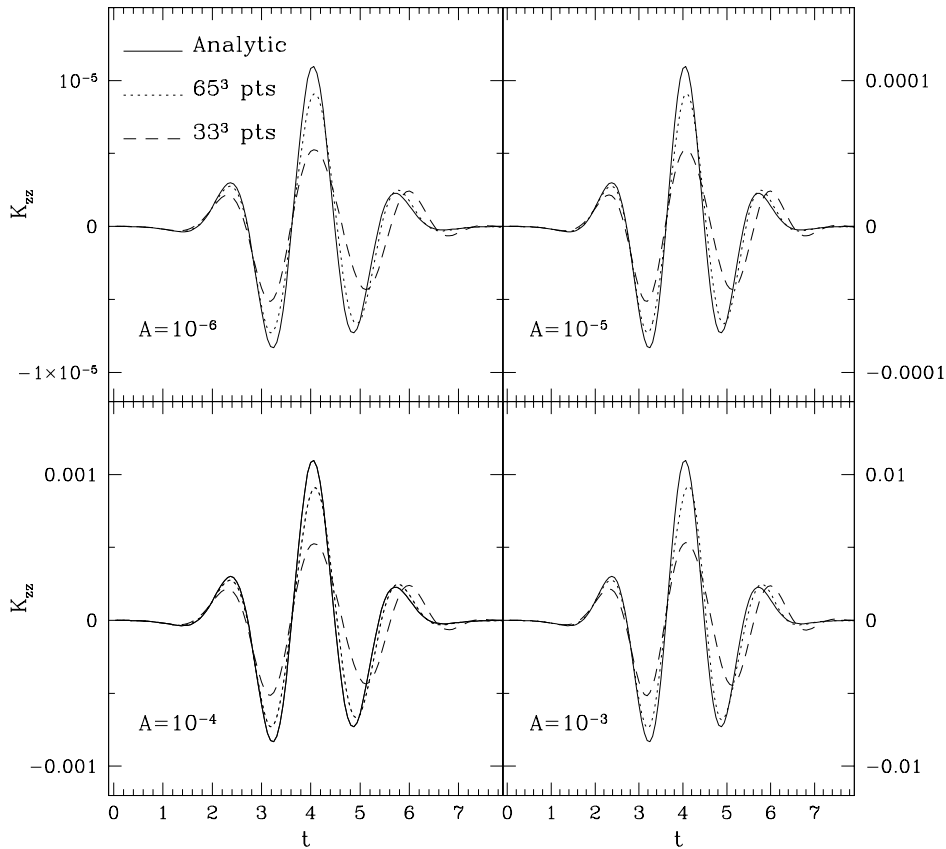}
\end{center}
\caption[fig14]{\label{fig14}
	Timeseries of the reconstructed values of $K_{zz}$ as measured
	at the gridpoint (4,0,0). The four different diagrams refer to
	the four different amplitudes used for the initial wave
	amplitude (i.e. $A = 10^{-6}-10^{-3}$, cf. Fig. \ref{fig4},
	Fig. \ref{fig13}). The integration scheme used is
	Crank-Nicholson.}
\end{figure}

	Results of these tests are presented in Figs. \ref{fig13} for
the extraction of gravitational waves and in Fig. \ref{fig14} for the
imposition of outer boundary conditions (for these tests we have gone
back to the less computationally expensive $\ell=2$, $m=0$ initial
wave packet). Fig. \ref{fig13}, in particular, shows the timeseries of
the multipole amplitude $(a_+)_{_{2 0}}$ extracted at a 2-sphere of
radius $r_{_E} = 1.0$ for amplitudes ranging from $A = 10^{-6}$ to $A
= 10^{-3}$. In is interesting to note how all diagrams are almost
perfectly identical but for the different scale used. Similar
considerations apply also for the boundary conditions imposed at the
outer boundary and shown for the ${zz}$ component of the extrinsic
curvature in Fig. \ref{fig14}

\section{Conclusion}

	We have investigated the properties of the Cauchy perturbative
method for matching gravitational data computed from a 3D Cauchy
solution of Einstein field equations. Studying the evolution of linear
and mildly nonlinear waves we have shown the ability of the
perturbative module to extract convergent gravitational wave
information at different locations within the 3D numerical grid
solving the nonlinear form of Einstein's equation. We have shown that,
given an extraction 2-sphere radius, a resolution can be found which
provides extraction and reinjection with the required accuracy.

	We have also discussed in detail a number of different
approaches to the problem of imposing outer boundary conditions.
Relying on the important advantage of being able to provide
information on the whole portion of the 3D numerical grid outside the
extraction 2-sphere, we have investigated Dirichlet boundary
conditions, perturbative Sommerfeld boundary conditions and
``perturbative blended'' boundary conditions. Each of these approaches
has been shown to provide convergent boundary conditions, but only the
latter provides stable evolutions. It has been recognized for sometime
that it is advantageous to make variable choices for numerical
relativity which separate those variables with dominantly wavelike
character from those dominated by static/stationary field moments
(\cite{e84,sp85,ae,nok87,sn95}). In the tests presented here, the
perturbative module provided a brute-force imposition of this
separation and, with suitable numerical implementation, enabled
long-term stability.

	These first successes of Cauchy perturbative matching method
motivate further work in this direction, both in the application of
our numerical module to fully nonlinear spacetimes and in the
extension of the mathematical apparatus to more general background
spacetimes.

\acknowledgments

	It is a pleasure to thank Roberto G\'omez, Jeff Winicour,
Thomas Baumgarte and Matt Choptuik for helpful discussions and
suggestions. We are also grateful to Greg Cook, Mijan Huq, Scott
Klasky and Mark Scheel for providing the interior evolution code. This
work was supported by the NSF Binary Black Hole Grand Challenge Grant
Nos. NSF PHY 93--18152, NSF PHY 93--10083, ASC 93--18152 (ARPA
supplemented) and by the NSF Grant AST 96--18524 to the University of
Illinois at Urbana-Champaign.  Computations were performed at NPAC
(Syracuse University) and at NCSA (University of Illinois at
Urbana-Champaign).

\appendix
\section{TOY MODEL FOR BLENDING BOUNDARY CONDITIONS}

	The very simplest model for investigating and interpreting the
effects of perturbative blending boundary conditions is provided by
linearized gravity waves in one dimension. The dynamical equations can
be written symbolically as

\begin{equation}
\label{gdot}
\dot{{\bf g}} \equiv \partial_t {\bf g} \sim -{\bf K} \ ,
\end{equation}

\begin{equation}
\dot{{\bf K}} \sim -{\bf g}'' \equiv -\partial^2_x {\bf g} \ ,
\end{equation}
where ${\bf g}$ and ${\bf K}$ are the fully nonlinear three-metric and
extrinsic curvature tensors respectively, whose numerical solution is
provided by the interior code. The blending condition on the extrinsic
curvature within the blending region can then be expressed as

\begin{equation}
\label{bbc}
{\bf K}_{_B} \equiv a(x){\bf K} + [1-a(x)]{\bf K}_{pert}
\end{equation}
where $a(x)$ is the ``blending function'', continuous in the range $x
\in [x_1, x_2]$ and defined so that $a(x_1)=1$ (i.e. at the inner edge
of the blending layer) and $a(x_2)=0$ (i.e. at the outer edge of the
blending layer). ${\bf K}_{pert}$ could be the perturbative solution
of Einstein's equations in the blending region (Note that the
following arguments will not depend on ${\bf K}_{pert}$ being a
perturbative solution. In fact, in G\'omez's first application of the
blending boundary conditions, ${\bf K}_{pert}$ was the value given by
an analytic solution \cite{rg97}.)

	Applying the boundary condition (\ref{bbc}) then yields

\begin{equation}
\dot{{\bf g}}_{_B} \sim -{\bf K}_{_B} \sim -a(x){\bf K} - 
	              [1-a(x)]{\bf K}_{pert} \ ,
\end{equation}
and therefore

\begin{equation}
\label{gddot}
\ddot{{\bf g}}_{_B} \sim a(x) {\bf g}'' - [1-a(x)]\dot{{\bf K}}_{pert}\ .
\end{equation}
In the simplest case in which the perturbative value for the extrinsic
curvature is zero, the result of the blending is then a wave equation
(in this case, in fact, ${\bf g}'' = {\bf g}_B''$) with a variable
phase propagation speed, $v_p^2(x) = a(x)$, which is unity at the
inner radius of the blending region and ``smoothly'' (in a discretized
sense) goes to zero at the outer edge of the blending region:

\begin{equation}
\ddot{{\bf g}}_{_B} \sim a(x) {\bf g}'' \ .
\end{equation}

	A less trivial and more interesting case is the one in which
an outgoing Sommerfeld condition is imposed within the blending
region. In this case:

\begin{equation}
{\bf g}' \sim - \dot{\bf g} \sim {\bf K}_{pert} \ ,
\end{equation}
and the ``blended'' equivalents of equations (\ref{gdot})
and (\ref{gddot}) are 

\begin{equation}
\dot{{\bf g}}_{_B} = -a(x){\bf K} - [1-a(x)]{\bf g}' \ ,
\end{equation}

\begin{equation}
\label{ngddot}
\ddot{{\bf g}}_{_B} \sim -a(x)\dot{{\bf K}} - [1-a(x)]\dot{{\bf g}}'
                    \sim  a(x) {\bf g}'' - [1-a(x)] \dot{{\bf g}}' \ .
\end{equation}

	Consider now an outgoing packet so that ${\bf g} \propto {\rm
const} \times e^{i(\omega t - kx)}$ and similarly for ${\bf
g}_{_B}$. The dispersion relation following from (\ref{ngddot}) will
be then

\begin{equation}
\omega^{2} - (1-a)k \omega + ak^{2} = 0 \ ,
\end{equation}
whose solutions are

\begin{equation}
\label{pv}
v_p(x) = \frac{\omega(x)}{k(x)} = \frac{[1-a(x) \pm (1+a(x))]}{2}
	   		        = (1, -a(x)) \ .
\end{equation} 

	 As a result, the wave packet will have unit outgoing and
ingoing phase velocities at the inner edge of the blending region,
decreasingly smaller ingoing phase velocities in the blending region
and only outgoing phase velocity at the outer edge of the blending
region and outside of it. 

	Finally, it can be seen that the blending approach, especially
the blending to the Sommerfeld condition, just described, has a close
relation to the techniques described in reference \cite{io81}.
However, for the strongly nonlinear black hole simulations, the more
manageable approach of blending with {\it analytic solutions}, as in
the first example above, has been successful in some cases
\cite{mh98}. This method is computationally much simpler than
explicitly modifying the equations.

\end{document}